\documentclass[aps,prb,twocolumn,groupedaddress]{revtex4}

\usepackage{graphics} 
\usepackage{graphicx}
\usepackage{dcolumn}
\usepackage{bm}
\everymath{\displaystyle}
\usepackage{amsmath}

\setcitestyle{super}

\begin{document}

\title{Positive Quantum Magnetoresistance in Tilted Magnetic Field.} 

\author{William Mayer}
\author{Areg Ghazaryan}
\author{Pouyan Ghaemi}
\author{Sergey Vitkalov}
\email[Corresponding author: ]{svitkalov@ccny.cuny.edu}
\affiliation{Physics Department, City College of the City University of New York, New York 10031, USA}
\author{A. A. Bykov}
\affiliation{A.V.Rzhanov Institute of Semiconductor Physics, Novosibirsk 630090, Russia}
\affiliation{Physics Department, Novosibirsk State University, Novosibirsk 630090, Russia}

\date{\today}

\begin{abstract} 

Transport properties  of highly mobile 2D electrons are studied in  symmetric GaAs  quantum wells placed in  titled magnetic fields. Quantum positive magnetoresistance (QPMR) is observed in  magnetic fields perpendicular to the 2D layer.  Application of  in-plane magnetic  field produces a dramatic  decrease of the  QPMR. This decrease correlates strongly with the reduction of the amplitude of Shubnikov de Haas resistance oscillations  due to modification of the electron spectrum via enhanced Zeeman splitting. Surprisingly no quantization of the spectrum is detected when  the Zeeman energy exceeds the half of the cyclotron energy suggesting an abrupt transformation of the electron dynamics.  Observed angular evolution of QPMR implies  $strong$  mixing between  spin subbands. Theoretical estimations  indicate that in the presence of spin-orbital interaction the elastic impurity scattering  provides  significant contribution to the spin mixing  in GaAs quantum wells at high filling factors.
\end{abstract}
 

\maketitle

\section{Introduction}

The orbital quantization of electron motion in magnetic fields generates a great variety of  fascinating transport phenomena observed in condensed materials.  Shubnikov-de Haas (SdH) resistance oscillations\cite{shoenberg1984} and   Quantum Hall Effect (QHE)\cite{qhe} are  famous examples. Spin degrees of freedom enrich  the electron response.\cite{ando1982,dyakonov_book} 
In two dimensional (2D) electron systems the orbital quantization is due to the component of the magnetic field, $B_\perp$, which is  perpendicular  to 2D layer\cite{fowler1966} whereas the spin degrees of freedom are affected mostly by the total magnetic field, $B$.\cite{fang1968}  An increase of the in-plane magnetic field produces, thus, an enhancement of the spin splitting (Zeeman effect), $\Delta_Z =\mu g B$, with respect to the cyclotron energy, $\Delta_C=\hbar \omega_c$. Here $\mu$ is Bohr magneton, $g$ is $g$-factor, $\omega_c=eB_\perp/m=eB cos(\alpha)/m$ is cyclotron frequency,  $m$ is electron effective mass and $\alpha$ is the angle between magnetic field $\vec B$ and the normal $\vec n$ to 2D layer.   At a critical  angle $\alpha_c$ corresponding to the condition: 
\begin{align}
\Delta_Z=\frac{\Delta_C}{2} \Leftrightarrow  cos(\alpha_c)=\frac{gm}{m_0},
\label{node}
\end{align}   
where parameter $m_0$ is mass of free electron, quantum levels are equally separated by $\hbar \omega_c/2$ and the amplitude of the fundamental harmonic of SdH oscillations, $A_{SdH}$,  is zero.   This property  is the basis of  a powerful  transport method (coincidence method) for  the study of the spin degrees of freedom of  2D electrons.\cite{ando1982,fang1968}    

In GaAs quantum wells the critical angle $\alpha_c$ is large: $\alpha_c \approx$ 85-87$^o$ due to a small effective electron mass. \cite{nicholas1988,leadley1998,piot2007,zudov2012} At low temperatures, $kT \ll \Delta_c$, the  coincidence method yields $g$-factor, which is considerably larger than the  one obtained from electron spin resonance.\cite{stein1983,kukush2011} Even stronger  spin gap is found in measurements of the  activation  temperature dependence of the magnetoresistance.\cite{nicholas1988,leadley1998}  The enhancement of the spin splitting  is  attributed to effects of electron-electron interaction of 2D electrons.\cite{ando1982} At low temperatures and high filling factors the spin splitting is found to be proportional to $B_\perp$ \cite{usher1990,dolg1997,leadley1998}, which agrees with  theoretical evaluations of  the contribution of the $e-e$ interaction to the spin gap, when only one quantum level is partially filled\cite{aleiner1995}.  

The  enhancement of the spin splitting is found above a sample dependent critical magnetic field $B_c$.\cite{group,leadley1998,piot2007} This effect has been attributed to the suppression of the contributions of the $e-e$ interaction to the spin splitting by a static disorder.\cite{fogler1995}  With an  increase of the temperature from mK range to few Kelvin the g-factor enhancement ($B_c$) is found to be decreasing (increasing) considerably, which is attributed to a reduction of the contribution of the $e-e$ interaction to the spin splitting due to thermal fluctuations. \cite{leadley1998}    

At high temperatures, $kT\gg \Delta_C, \Delta_Z$  there are many partially populated Landau levels  participating in transport and one may expect a quantitatively different value of the $e-e$ enhanced spin splitting in comparison with the one at $kT\ll \Delta_C, \Delta_Z$. We note that the spin splitting has not been  investigated experimentally  in the quantized  spectrum at high temperatures since the coincidence method  relies on SdH oscillations, which are absent (exponentially suppressed) in the high temperature  regime\cite{shoenberg1984}.  Recent developments  \cite{dietrich2012} open a possibility to study spin effects  in electron systems with quantized spectrum at  high temperatures: $kT\gg \Delta_C, \Delta_Z$. 

This paper presents an experimental  investigation of the Quantum Positive MagnetoResistance (QPMR) at high temperatures $kT \gg \Delta_C, \Delta_Z$ and SdH resistance oscillations in GaAs symmetric quantum wells placed in tilted quantizing magnetic fields.  The experiments indicate that angular variations of the QPMR and the SdH amplitude  strongly correlate yielding essentially the same $g$-factor: $g\approx$0.97$\pm$0.08. This  $g$-factor value is close to one  obtained  in  experiments   done at much lower temperatures.\cite{nicholas1988,leadley1998,piot2007,zudov2012} 

 At a fixed $B_\perp$$>$0.3T  the angular evolution of QPMR demonstrates a resistance maximum at $\alpha \approx$62$^o$, revealing  an unexpected decrease of the spin splitting with the in-plane magnetic fields while the overall angular evolution of QPMR  demonstrates   $B/B_\perp$ scaling at $\Delta_Z/\Delta_C$$<$1/2 ($\alpha$$<$86$^o$).  At $\alpha >$ 86$^o$ the QPMR does not return as expected indicating an absence of the quantized  electron spectrum in the  high temperature and large parallel field regime.  A  complementary study of quantal heating\cite{vitkalov2009,mamani2009,romero2008} at different angles  confirms this observation.     

In contrast to SdH oscillations the angular evolution of QPMR implies a significant $mixing$ between spin-up and spin-down subbands due to quadratic dependence of the conductivity on DOS (see Eq.(\ref{cond2})). When the spin and momentum of the electrons are independent, the non-magnetic impurities can not mix the electronic states with opposite spins. On the other hand in presence of spin-orbit coupling, the spin and momentum of electrons are not independent.  In contrast to the Zeeman splitting the spin-orbit  interaction  depends on the energy (velocity) of electrons  and does not  decrease at small magnetic fields.\cite{rashba1960,bychkov1984}  As we show below, even at a small spin-orbit coupling local non-magnetic impurities may  lead to a scattering between different  subsets of quantum levels leading to  the spin mixing  at high filling factors. 
 
\section{Experimental Setup}

Studied GaAs quantum wells were grown by molecular beam epitaxy on a semi-insulating (001) GaAs substrate. The material was fabricated from a selectively doped GaAs single quantum well of width $d=$13 nm sandwiched between AlAs/GaAs superlattice barriers.  The studied samples were etched in the shape of a Hall bar. The width and the length of the measured part of the samples are $W=50\mu$m and $L=250\mu$m. AuGe eutectic was used to provide electric contacts to the 2D electron gas. Two samples were studied at temperature 4.2 Kelvin in magnetic fields up to 9 Tesla applied $in$-$situ$ at different angle $\alpha$  relative  to the normal to 2D layers and  perpendicular to the applied current.  The angle $\alpha$ has been evaluated using Hall voltage $V_H = B_\perp/(en_T)$, which is proportional to the perpendicular component, $B_\perp=B\cdot cos(\alpha)$, of the total magnetic field $B$.     The total electron density of samples, $n_T\approx 8.6\times 10^{11} cm^{-2}$, was evaluated from the Hall measurements taken at $\alpha$=0$^0$ in classically strong magnetic fields \cite{ziman}. An average electron mobility $\mu \approx 1.6 \times 10^6 cm^2/Vs$  was obtained from $n_T$ and the zero-field resistivity.  Sample resistance was measured using the four-point probe method. We applied a 133 Hz $ac$ excitation $I_{ac}$=1$\mu$A  through the current contacts and measured the longitudinal (in the direction of the electric current, $x$-direction) and Hall $ac$ (along $y$-direction) voltages ($V^{ac}_{xx}$ and $V^{ac}_H$) using two lockin amplifiers with 10M$\Omega$ input impedances.  The measurements were done in the linear regime in which the voltages are proportional to the applied current.

\section{Results and Discussion}

\begin{figure}[t]
\vskip -0.4cm
\includegraphics[width=\columnwidth]{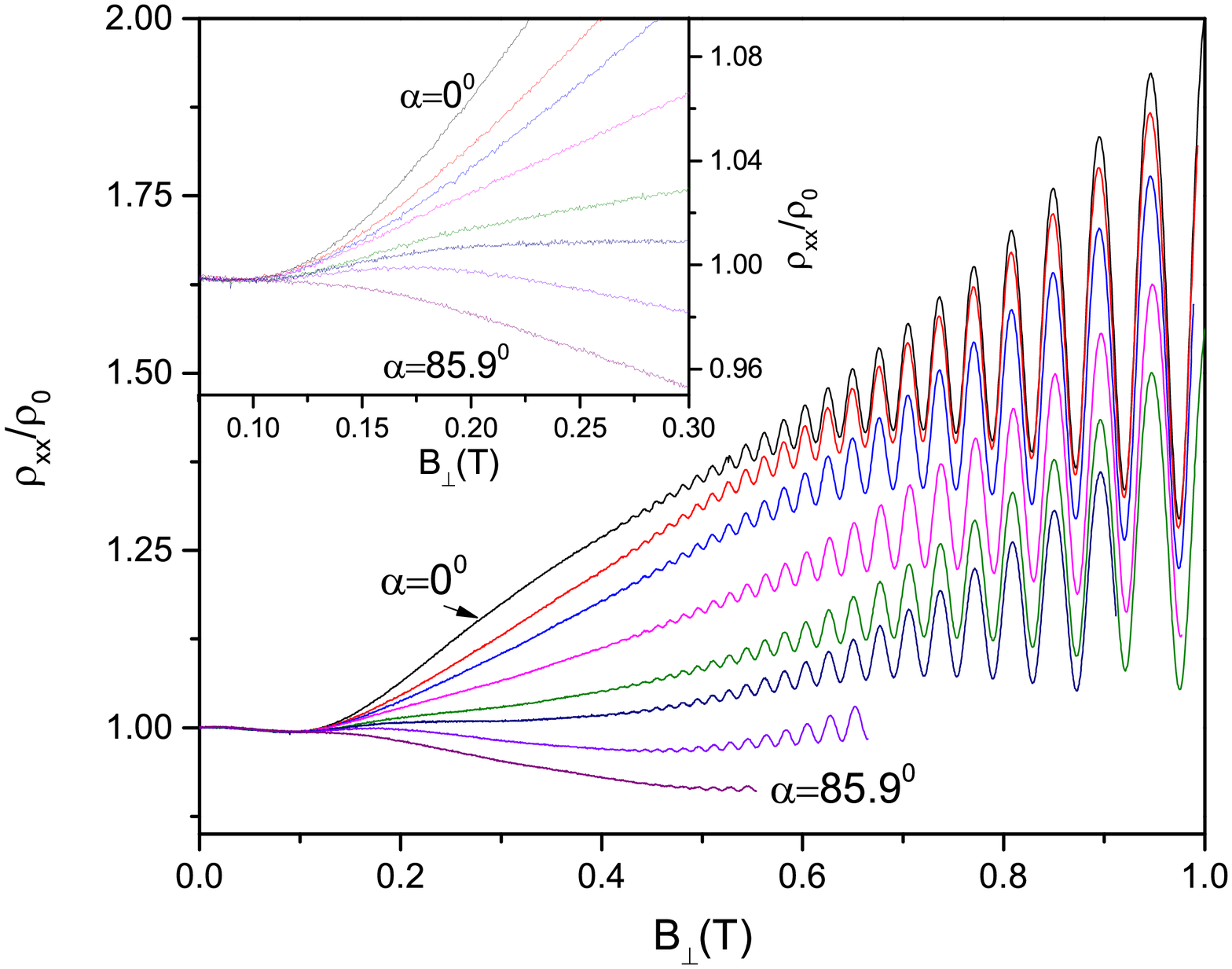}
\caption{(Color online) Dependence of the longitudinal resistance $\rho_{xx}$ on the magnetic field perpendicular to the 2D sample obtained at different angles $\alpha$ between the total magnetic field $\vec B$ and the normal to the 2D layer. From the top curve to the bottom one angles are 0, 76.2, 78.6, 81.2, 82.6 83.13, 84, and 85.9 degrees.   The insert enlarges the area at small magnetic fields indicating that the dependencies at different angles diverge from approximately the same magnetic field $B^*\approx$0.11T corresponding to the beginning of Landau quantization of the electron spectrum at $\alpha$=0$^0$.  }
\label{qpmr}
\end{figure}

Figure \ref{qpmr} presents the dissipative resistivity, $\rho_{xx}(B_\perp)$,   at different angles $\alpha$ between the magnetic field, $\vec B$, and the normal to the  2D layer, $\vec n$. In perpendicular magnetic fields below 0.11 T the resistance  is nearly (within $\sim$0.6\%) independent on  $B_\perp$.  This is  the regime of  classical (Drude) magnetoresistance, which is expected to be independent on $B_\perp$.\cite{ziman} 

 At $\alpha=0^0$ and  $B_\perp>$0.11 T the magnetoresistance demonstrates a steep (exponential in $1/B_\perp$) monotonic increase combined with  SdH oscillations  in $B_\perp>$ 0.45 T.   This  increase is attributed\cite{dietrich2012}  to the quantum positive magnetoresistance (QPMR) due to Landau quantization.\cite{vavilov2004}   At angles $\alpha <$ 65$^o$ and $B_\perp >$0.33 T the magnetoresistance exhibits an additional few percent increase with the angle (not shown).   At $\alpha >$65$^0$ the QPMR decreases significantly with the angle.  Figure \ref{qpmr} demonstrates this decrease for  angles  between 76.2 and 85.9 degrees.   The insert to the figure shows that the angular variation of QPMR are approximately uniform with $B_\perp$ and starts at  the $same$  perpendicular magnetic field $B^*\approx$0.11 T, which  separates the classical and quantum regimes of  electron transport\cite{vavilov2004,dietrich2012}.   The later indicates that the Landau level width (or the quantum scattering time $\tau_q$) is nearly independent of angle $\alpha$. This is confirmed by  more detailed comparison  (see Fig.\ref{qpmr2}(d)).  At  angles $\alpha >$86$^0$ the QPMR demonstrates a weak recovery (not shown),  which is discussed later.  

At $B_\perp>$0.45T Fig.\ref{qpmr} shows SdH oscillations.  In contrast to  QPMR the angular evolution of SdH oscillations have been intensively studied.\cite{nicholas1988,leadley1998,zudov2012}  Presented experiments show $strong$ correlation between angular evolutions of SdH oscillations and QPMR. Before the detail discussion and comparison of  the  dependencies, we present a model, which captures the strong angular correlations of these two phenomena.

\subsection{Model of SdH oscillations and Quantum Positive Magnetoresistance}

A  microscopic description of both SdH oscillations and QPMR in perpendicular magnetic fields (at $\alpha=$0$^o$) is presented in paper Ref.[\cite{vavilov2004}] neglecting any spin related effects in particular the Zeeman term.  As indicated in the  "Introduction" the account of the Zeeman splitting for SdH oscillations is a developed procedure\cite{fang1968,ando1982,leadley1998}. In contrast  the spin related effects in the QPMR  have not been studied yet. 

Below we present a  model, which utilizes the similarity of QPMR and Magneto-InterSubband (MISO) resistance oscillations.\cite{coleridge1990,leadley1992,raikh1994,raichev2008,mayer2016}  The model considers two subbands with the energy spectrum evolving in accordance with  Landau quantization and splitted predominantly by  Zeeman effect.\cite{fang1968}    A scattering assisted mixing between different  subbands is $postulated$ to provide the observed correlation between the angular evolutions of SdH oscillations and QPMR. Within the presented model the absence of the scattering  between  subbands would lead to the absence of an angular evolution of the QPMR associated with the Zeeman effect in contrast to the angular dependence of SdH oscillations.  The origin of the mixing requires further investigations.  A  mixing between different spin subbands have been reported in Si-MOSFETs.\cite{vitkalov2000} The experiments show a sizable  contributions of the product the spin-up and spin-down density of states to SdH resistance oscillations. Furthermore investigations of the resistivity tensor in tilted magnetic fields have revealed an independence of  the Hall coefficient  on  the  spin subband populations while the  electron mobility in each spin subband was substantially affected by the in-plane magnetic field\cite{vitkalov_hall}. This behavior  has been interpreted by  a mixing between spin subbands due to an electron-electron interaction.\cite{vitkalov_hall2}           We note also that in the presence of a spin-orbit coupling, different subbands could be mixed by a local impurity scattering. An investigation of this possibility  is presented in the section "Spin orbit interaction and QPMR". 

In the simplest case of small quantizing magnetic fields  $\omega_c \tau_q < 1$ the main contribution to both SdH oscillation and QPMR comes from the fundamental harmonic of quantum oscillations of the density of states (DOS) corresponding to spin-up and spin-down subbands. The total DOS, $\nu(\epsilon)$, reads\cite{ando1982}:    
\begin{equation}
\begin{split}
\nu(\epsilon)\!&=\!\nu_0\!\left[\!1\!-\!\delta cos\!\left(\!\frac{2\pi (\epsilon\!-\!\Delta_Z/2)}{\hbar \omega_c}\!\right)\!-\!\delta cos\!\left(\!\frac{2\pi (\epsilon\!+\!\Delta_Z/2)}{\hbar \omega_c}\!\right)\!\right] \\ 
&= \nu_0\left[1-2\delta cos\left(\frac{2\pi \epsilon}{\hbar \omega_c}\right)cos\left(\frac{\pi \Delta_Z}{\hbar \omega_c}\right)\right]
\end{split}
\label{dos}
\end{equation}
where $\delta=exp(-\pi/\omega_c \tau_q)$ is Dingle factor, $\nu_0$ is the total  DOS at  zero magnetic field and $\tau_q$ is the quantum scattering time, which is considered to be the same in both spin subbands.
 
The 2D conductivity $\sigma$ is obtained from the following relation:
\begin{equation}
\sigma(B)=\int d\epsilon \sigma(\epsilon)\left(-\frac{\partial f}{\partial \epsilon}\right)=\langle \sigma(\epsilon) \rangle
\label{cond}
\end{equation}  
The integral is  an average of the conductivity $\sigma(\epsilon)$ taken essentially for  energies  $\epsilon$  inside the  temperature interval $kT$ near Fermi energy, where $f(\epsilon)$ is the electron distribution function at the temperature $T$. \cite{ando1982} The brackets represent this integral below.   

The following expression  approximates the conductivity $\sigma(\epsilon)$ at small quantizing magnetic fields:
\begin{equation}
\sigma(\epsilon,B_\perp,\Delta_Z)=\sigma_D(B_\perp)\tilde{\nu}(\epsilon,B_\perp,\Delta_Z)^2
\label{cond2}
\end{equation}
where $\sigma_D(B_\perp)$ is Drude conductivity in magnetic field $B_\perp$ \cite{ziman} and $\tilde{\nu}(\epsilon)=\nu(\epsilon)/\nu_0$ is normalized total density of states. The main assumption of this model is utilized in Eq.(\ref{cond2}). Namely  the impurity scattering between the spin-up and spin-down subbands is considered to be comparable with the impurity scattering within a spin subband, when the energies of the spin sectors are the same.   In other words a spin up (spin-down) electron has equal probability to scatter into a spin-up or spin-down quantum state. 

The proportionality of the conductivity $\sigma(\epsilon)$ to the square of the normalized density of states is due to two factors. One factor takes into account the number of available conducting states (parallel channels) at energy $\epsilon$, which is proportional to the density of states. The second factor takes into account that the dissipative conductivity in crossed electric and magnetic fields is proportional to the electron scattering rate\cite{ziman}. At low temperatures the scattering  is dominated by the elastic impurities  making  the  rate proportional to the density of final states at the same energy $\epsilon$.\cite{ziman,gantmakher1987}  The quadratic dependence of the conductivity on the density of state results in the factor 4 in Eq.(\ref{SdH}), which is found to be in good quantitative agreement with the amplitude of SdH oscillations  shown in Fig.\ref{sdh}. Furthermore the quadratic dependence on the density of states yields both  QPMR and its strong correlation with SdH oscillations observed in presented experiments. 

The Eq.(\ref{cond2}) is similar to the Eq.(5) of Ref.[35], which was used for the conductivity in the perpendicular magnetic fields neglecting both the Zeeman splitting and spin-orbital effects. In this case the energy spectrum of spin-up and spin-down electrons are  the same and the normalized DOS for each spin subband coincides with  the normalized total DOS, $\tilde\nu(\epsilon)$. For two independent spin subbands the total conductivity is the sum of two terms: $\sigma_{ind}=\sigma^++\sigma^-$, where $\sigma^\pm=(\sigma_D/2)\tilde \nu(\epsilon)^2$. The factor $1/2$ takes into account that the electron density  in each subbands is half the total density.  Thus at $\Delta_Z$=0 the total conductivity of two subbands does not depend on the intersubband scattering: $\sigma_{ind}=\sigma(\Delta_Z=0)$.     At finite $\Delta_Z$  the intersubband scattering affects the conductivity.  

A substitution of  Eq.(\ref{cond2}) and Eq.(\ref{dos}) into Eq.(\ref{cond}) yields two  additional terms  to the Drude conductivity: $\sigma-\sigma_D=\Delta\sigma_{SdH}+\Delta\sigma_{QPMR}$. The first term is proportional to Dingle factor $\delta$ and describes SdH oscillations. It reads:   
\begin{equation}
\begin{split}
\frac{\Delta\sigma_{SdH}}{\sigma_D}&=-4\delta \langle \cos\left(\frac{2\pi \epsilon}{\hbar \omega_c}\right)\rangle \cos\left(\frac{\pi \Delta_Z}{\hbar \omega_c}\right ) \\ 
&=-4\delta A(T)\cos\left(\frac{2\pi\epsilon_F}{\hbar\omega_c}\right) \cos\left(\frac{\pi \Delta_Z}{\hbar \omega_c}\right),
\end{split}
\label{SdH}
\end{equation}
where  $\epsilon_F$ is Fermi energy and $A(T)=\frac{(2\pi^2kT/\hbar\omega_c)}{\sinh(2\pi^2kT/\hbar\omega_c)}$ is SdH temperature factor.\cite{shoenberg1984}

The second term is proportional to the square of the Dingle factor and  describes variations of the conductivity due to QPMR. It reads:   
\begin{equation}
\begin{split}
\frac{\Delta\sigma_{QPMR}}{\sigma_D}&=4\delta^2 \langle \cos^2\left(\frac{2\pi \epsilon}{\hbar \omega_c}\right)\rangle \cos^2\left(\frac{\pi \Delta_Z}{\hbar \omega_c}\right ) \\ 
&=\delta^2 \left[1+\cos\left(\frac{2\pi \Delta_Z}{\hbar \omega_c}\right)\right],
\end{split}
\label{QPMR}
\end{equation}
In Eq.(\ref{QPMR}) the  exponentially small temperature dependent term is neglected. At $\Delta_Z$=0 Eq.(\ref{QPMR}) reproduces QPMR in perpendicular magnetic fields.\cite{vavilov2004,dietrich2012}

Eq.(\ref{SdH}) and Eq.(\ref{QPMR}) indicate the strong angular correlation between the amplitude of SdH oscillations and the QPMR. In particular the SdH amplitude is proportional to $\cos(\pi\Delta_Z/\hbar\omega_c)$ and is zero at $\Delta_Z=\hbar\omega_c/2$ in agreement with  Eq.(\ref{node}), while the QPMR is proportional to (1+$\cos(2\pi\Delta_Z/\hbar\omega_c)$ and is zero too at $\Delta_Z=\hbar\omega_c/2$.  In the next sections we compare experimental results with Eq.(\ref{SdH}) and Eq.(\ref{QPMR}).

\subsection{Shubnikov de Haas oscillations}

\begin{figure}[t]
\vskip 0 cm
\includegraphics[width=\columnwidth]{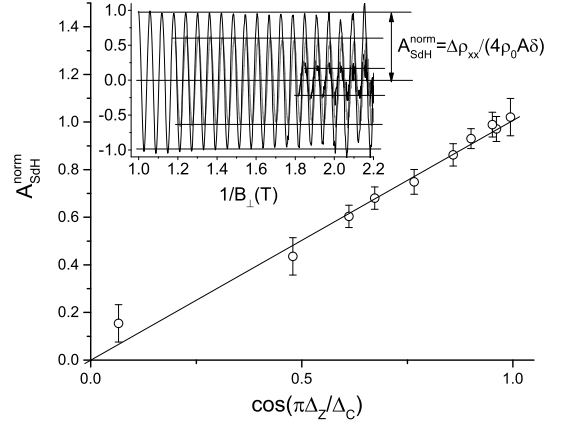}
\caption{Dependence of normalized amplitude of SdH oscillations, $A_{SdH}^{norm}=A_{SdH}/4\delta A(T)$ on $\cos(\pi\Delta_Z/\Delta_C)$ with $g$=0.97. The dependence corresponds to Eq.(\ref{SdH}) relating the angular evolution of SdH amplitude to the angular variation of the ratio between Zeeman and cyclotron energies: $\Delta_Z/\Delta_C= m g /(2m_0\cos\alpha)$. The insert presents normalized SdH resistance oscillations in reciprocal magnetic fields at angles $\alpha$: 67.6, 83.3 and 85.9 degrees.  }
\label{sdh}
\end{figure}

In quantizing magnetic fields  $\omega_c\tau_{tr} \gg$1, where $\tau_{tr}$ is the transport scattering time. At this condition resistivity is $\rho_{xx}=\sigma[\rho_{xy}]^2$ and $\rho_{xx}(B_\perp)/\rho_0=\sigma(B_\perp)/\sigma_D(B_\perp)$, where $\rho_0$ is Drude resistivity.\cite{ziman}   Therefore in accordance with Eq.(\ref{SdH}) the amplitude of SdH oscillations of the normalized resistivity, $\Delta\rho_{SdH}/\rho_0$, is $A_{SdH}=4\delta A(T)\cos(\pi\Delta_Z/\hbar\omega_c)$ and the normalized SdH amplitude is  $A_{SdH}^{norm}=A_{SdH}/(4\delta A(T))=\cos(\pi\Delta_Z/\hbar\omega_c)$.  To extract the normalized  amplitude $A_{SdH}^{norm}$, the  SdH resistance oscillations  shown in Fig.\ref{qpmr}  were separated from the monotonic background using a low frequency filtering.\cite{mayer2016}   The separated SdH oscillations were then divided  by the factor $4\rho_0\delta(B_\perp,\tau_q)  A(T)$.   By a variation of the quantum scattering time $\tau_q$ in the Dingle factor $\delta$ quantum oscillations with the amplitude, $A_{SdH}^{norm}$, independent on the magnetic field, $B_\perp$,   are obtained. The later indicates that the ratio of the Zeeman energy, $\Delta_Z$ to the cyclotron energy, $\Delta_C=\hbar \omega_c$ is a constant at fixed angle $\alpha$ in the SdH regime.  The insert to Fig.\ref{sdh} shows the independence of the normalized SdH amplitude, $A_{SdH}^{norm}$, on the reciprocal magnetic fields at different angles $\alpha$. 

Figure \ref{sdh} presents the angular dependence of the normalized SdH amplitude $A_{SdH}^{norm}$.  
 We note that the value of the SdH amplitude agrees quantitatively with the one expected from Eq.(\ref{SdH}). The dependence  is plotted versus $\cos(\pi\Delta_Z/\hbar\omega_c)=\cos(\pi m g /(2m_0\cos\alpha))$. The g-factor is used as a scaling  parameter for $x$-axes of the plot to provide the linear dependence between  $A_{SdH}^{norm}$ and $\cos(\pi\Delta_Z/\hbar\omega_c)$. The obtained value of g-factor $g$=0.97$\pm$0.08 corresponds to the critical angle $\alpha_c$=86.3$^o\pm$0.3$^o$ (see Eq.(\ref{node})) and is in a good agreement with existing experiments.\cite{nicholas1988,leadley1998,piot2007,zudov2012}  Thus the angular evolution of SdH oscillations agrees  with both Eq.(\ref{SdH}) and existing experiments.   We note that the  strong enhancement of the  g-factor  obtained in the present experiments in the  high  temperature regime  is intriguing, since the enhancement should degrade with temperature increase in the low temperature domain.\cite{leadley1998} 
 
The obtained quantum scattering rate, $1/\tau_q$, is presented in Fig.\ref{qpmr2}(d). The scattering rate is found to be independent on the angle $\alpha$: $1/\tau_q^{SdH}\approx 300 \pm100$GHz and agrees with the one obtained using the QPMR described in  next section.  
 
\subsection{Quantum Positive Magnetoresistance.}

\begin{figure}[t]
\vskip 0 cm
\includegraphics[width=\columnwidth]{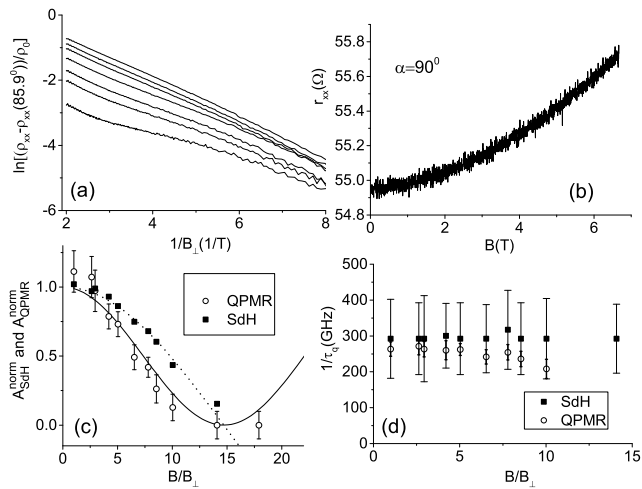}
\caption{(a) Dependence of the difference between normalized resistivity at an angle $\alpha$ and the normalized resistivity at $\alpha $=85.9$^0$$\approx$$ \alpha_c$:  $[\rho_{xx}-\rho_{xx}(85.9^0)]/\rho_0$ on the  reciprocal magnetic field. From the top curve to the bottom one corresponding angles $\alpha$ are  70.1, 76.2, 78.6, 81.2, 82.6, 83.3 and 84.3 degrees; (b) magnetoresistance at $\alpha$=90$^0$;  (c) Dependence of the normalized QPMR and SdH  amplitudes on the ratio between total and perpendicular magnetic fields. The solid (dotted) line presents the normalized QPMR (SdH) amplitude: $A_{QPMR}^{norm}=(1+cos(2\pi\Delta_Z/\hbar\omega_c))/2$ ($A_{SdH}^{norm}$) obtained from Eq.(\ref{QPMR}) (Eq.(\ref{SdH}))  using g-factor $g$=0.97;   (d)  Dependence of the quantum scattering rate on the ratio $B/B_\perp$ obtained from the analysis SdH oscillations and the exponential decrease of the QPMR magnitude  with $1/B_{\perp}$ expected from Eq.(\ref{QPMR}). }
\label{qpmr2}
\end{figure}

In accordance with Eq.(\ref{QPMR}) the magnitude of the quantum magnetoresistance decreases exponentially with the reciprocal magnetic field, $1/B_\perp$, due to the exponential decrease of Dingle factor $\delta$.  Below we explore this property of the QPMR to extract the quantum scattering rate $1/\tau_q$ and the normalized QPMR amplitude  $A_{QPMR}^{norm}=(1+\cos(2\pi\Delta_Z/\Delta_c))/2$.  In the vicinity of the critical angle $\alpha_c$ the magnitude of the QPMR is expected to be very small and the magnetoresistance should be mostly driven by other mechanisms.\cite{baskin1978,fogler1997,polyakov2001}   In particular Fig.\ref{qpmr2}(b) presents the magnetoresistance at angle $\alpha $=90$^0$ at which only in-plane  magnetic fields is applied. The resistance demonstrates a weak (within 2\%) parabolic increase with the in-plane magnetic field.   The small in-plane magnetoresistance affects weakly  the curves presented in Fig.\ref{qpmr} and can be taken into account assuming its independence on the angle $\alpha$.      Below we assume  that all  mechanisms leading to  negative magnetoresistance in the vicinity of the critical angle are independent on the angle $\alpha$ and controlled by $B_\perp$ and $B_\parallel$ independently.  
 
Within this assumption the difference between magnetoresistance at an angle $\alpha$ and the magnetoresistance at  the critical angle $\alpha_c$ captures the main effect of the angular variations of the electron spectrum  on the electron transport described by Eq.(\ref{QPMR}). Figure \ref{qpmr2}(a) presents the dependence of the difference between the resistivity $\rho_{xx}(\alpha)$ and $\rho_{xx}(85.9^0\approx \alpha_c)$ normalized to the Drude resistivity $\rho_0$ on the reciprocal magnetic field, $1/B_\perp$ taken at different angles.  At small magnetic fields, $B_\perp$, the dependences demonstrate an exponential decrease with $1/B_\perp$ in accord with Eq.(\ref{QPMR}) with the rate  depending weakly  on $\alpha$.  With an increase of the angle $\alpha$ the dependencies shift down indicating a decrease of the normalized  QPMR amplitude   $A_{QPMR}^{norm}$.  The presented resistance difference  takes into account the small variations of the resistivity with  the in-plane magnetic field shown in Fig.\ref{qpmr2}(b).  The applied correction to the resistivity affect very weakly (within the size of the symbols) the results presented in Fig.\ref{qpmr2}(c,d).        

Fig.\ref{qpmr2}(c) presents  the normalized  QPMR amplitude   $A_{QPMR}^{norm}$ and   SdH amplitude   $A_{SdH}^{norm}$ plotted vs $1/\cos\alpha=B/B_\perp$.  The normalized QPMR amplitude is obtained by the extrapolation of the  linear dependencies shown in Fig.\ref{qpmr2}(a) at high $1/B_\perp$ to the infinite $B_\perp$. The extracted normalized amplitude $A_{QPMR}^{norm}$ is presented by the open symbols. The solid line shows the amplitude  $A_{QPMR}^{norm}$ obtained from Eq.(\ref{QPMR}) using g-factor $g$=0.97. We note that there are no fitting parameters between the experiment (open symbols) and the Eq.(\ref{QPMR}) since the g-factor is obtained from the fitting of the angular dependence of the SdH amplitude. Shown in Fig.\ref{qpmr2}(c) comparison of two amplitudes indicates  strong angular correlations between SdH resistance oscillations and the quantum positive magnetoresistance. 

Fig.\ref{qpmr2}(d) presents  the quantum scattering rates  obtained from the analysis of SdH resistance oscillations (filled symbols) and QPMR (open symbols). In contrast to SdH resistance oscillations the analysis of the QPMR magnitude yields more accurate results for $\tau_q$ since QPMR does not depend on the temperature damping factor $A(T)$  and the response is mostly controlled by the Dingle factor only.   The quantum scattering rates extracted by two different methods  are found to be in a reasonable agreement indicating no significant variations of the electron lifetime $\tau_q$ with both the angle $\alpha$ and the applied magnetic fields at $\alpha<\alpha_c$.

Figures \ref{qpmr} and \ref{qpmr2}(a) demonstrate the evolution of the QPMR, which is obtained  at a fixed angle $\alpha$. At this condition both perpendicular and total magnetic fields are changing. As mentioned above the angular evolution of QPMR at small ($<$65$^0$) and large ($>\alpha_c$) angles demonstrates  additional features, which may required a  modification of the proposed description.  To get further insight into  the angular evolution of the QPMR, we have conducted measurements  at a fixed perpendicular magnetic field, $B_\perp$, while sweeping the in-plane magnetic field, $B_\parallel$.  At this condition the cyclotron energy is fixed and variations of the electron spectrum are related mostly  to spin degrees of freedom.

\begin{figure}[t]
\vskip 0 cm
\includegraphics[width=\columnwidth]{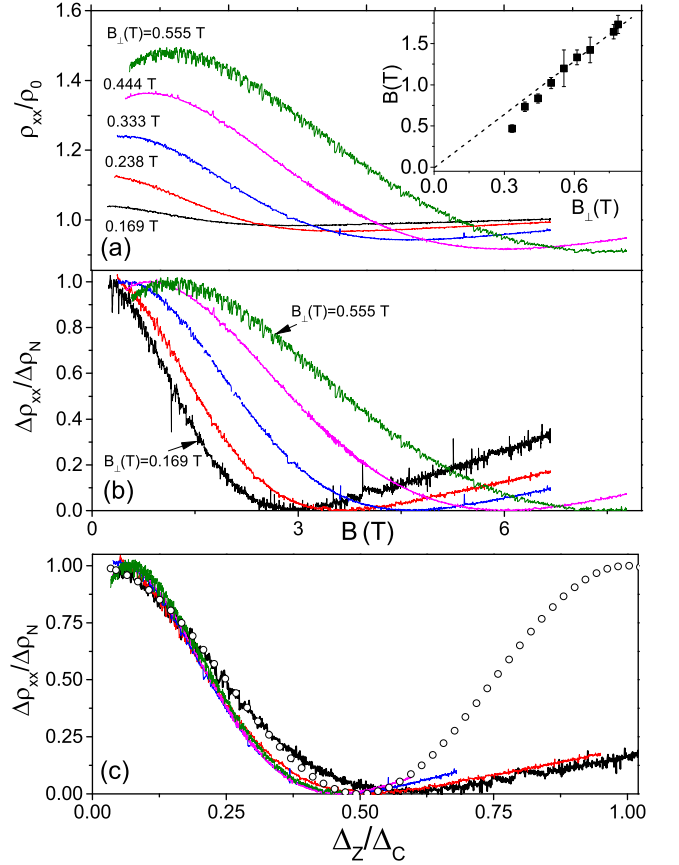}
\caption{(Color online) (a) Dependence of normalized resistivity on magnetic field at fixed $B_\perp$ as labeled. Insert shows position of the resistance maximum at different $B_\perp$; (b) Normalized variations of the resistivity shown in (a): $\Delta \rho_{xx}/\Delta\rho_N=(\rho_{xx}-\rho_{min})/(\rho_{max}-\rho_{min})$ vs $B$; (c) Normalized variations of the resistivity vs ratio of Zeeman and cyclotron energies, $\mu g B/\hbar\omega_c$, with g-factor $g$=0.97 obtained from  the angular variation of SdH oscillations. Open symbols present the normalized magnetoresistance expected from Eq.(\ref{QPMR}) with no fitting parameters.}
\label{qpmr3}
\end{figure}

Figure \ref{qpmr3}(a) presents  dependencies of the normalized resistivity on the total magnetic field  taken at the fixed  $B_\perp$ as labeled. In the agreement with the angular evolution shown in Fig.\ref{qpmr} the total magnetic field suppresses the quantum magnetoresistance  at a fixed $B_\perp$.    A stronger perpendicular magnetic field, $B_\perp$, requires a  stronger total magnetic field, $B$, to suppress the QPMR.   After the suppression the magnetoresistance demonstrates a weak increase with the  magnetic field, which is, however, much smaller than expected from Eq.(\ref{QPMR}).  Finally at $B_\perp>$0.3 T the magnetoresistance shows a maximum enhancing at higher $B_\perp$ that is also not explained by this  model. The insert to  Fig.\ref{qpmr3}(a) presents the position of the resistance maximum at different $B_\perp$ indicating that the maximum occurs at $\alpha_0 \approx$62$^0$.  

Figure \ref{qpmr3}(b) presents the  magnetic field dependencies of  a normalized resistance variation:  
$\Delta \rho_{xx}/\Delta\rho_N=(\rho_{xx}-\rho_{min})/(\rho_{max}-\rho_{min})$, where $\rho_{max}$ and $\rho_{min}$ are maximum and minimum values of the curves shown in (a).  The figure facilitates the comparison of the shape of the dependencies at different $B_\perp$. 

In accordance with the proposed model (see Eq.(\ref{QPMR})) at a fixed $B_\perp$ and a constant  quantum lifetime $\tau_q$ the Dingle factor is fixed and the evolution of the magnetoresistance is solely due to variations of the QPMR amplitude, $A_{QPMR}^{norm}=(1+cos(2\pi\Delta_Z/\hbar\omega_c))/2$.  If the g-factor is also a constant, then the Zeeman term, $\Delta_Z=\mu g B$,  is linearly proportional to the total magnetic field, $B$, and the QPMR amplitude depends only on the ratio $B/B_\perp$. Thus in this case the QPMR should scale with $B/B_\perp$.      

Figure \ref{qpmr3}(c) presents  the  normalized resistance variations,  
$\Delta \rho_{xx}/\Delta\rho_N$, shown in Fig.\ref{qpmr3}(b) plotted  against  the ratio between Zeeman and cyclotron energies: $\Delta_Z/\Delta_C=(m g /2m_0)(B/B_\perp)$, using the constant g-factor $g$=0.97 obtained from the angular dependence of the amplitude of  SdH oscillations.  Except the  curve taken at the smallest $B_\perp$=0.169 T  all other curves shown  in Fig.\ref{qpmr3}(a,b)  collapse  on a single dependence at $\Delta_Z/\Delta_C$ between 0.07 and 0.5.  The collapse indicates  $B/B_\perp$ scaling, which holds at  high $B_\perp$ in the studied system.

At $\Delta_Z/\Delta_C<$ 1/2 the scaled dependencies are quite close to the dependence expected from Eq.(\ref{QPMR}) and presented by the  open circles at $g$=0.97  in  Fig.\ref{qpmr3}(c) with no fitting parameters.  The dependence obtained at the smallest  $B_\perp$=0.169 T agrees  better with the model. We note that the  model  takes into account only fundamental harmonics of the electron spectrum and, thus, is valid only for overlapping Landau levels  at $\omega_c\tau_q <$1. At $B_\perp>$0.3T the Landau levels become separated at $\tau_q \approx$4 ps and an account of the higher harmonics of DOS may improve the agreement with the experiment at high $B_\perp$. In contrast the description of SdH oscillations is valid even  at higher $B_\perp$ since the contributions of the higher harmonics of DOS to the SdH amplitude are suppressed by the   temperature for presented $B_\perp$.\cite{shoenberg1984,vavilov2004}   We note also that the shift of the resistive variation at $B_\perp$=0.169T to a stronger $B$ ($\Delta_Z$) in Fig.\ref{qpmr3}(c) agrees with the reduction of the  enhanced g-factor by the disorder \cite{fogler1995,leadley1998}      

An unexpected feature of the dependences presented in Fig.\ref{qpmr3}  is  the resistance maximum emerging at high $B_\perp$.  In accordance with Eq.(\ref{QPMR}) the maximum occurs at $\Delta_Z$=0 and corresponds to the  alignment of  the quantum levels corresponding to spin-up and spin-down subbands. The presence of the maximum at a finite magnetic field, $B$, suggests that the magnitude of the Zeeman splitting, $\vert\Delta_Z(B_\perp)\vert$, decreases with the increase of the total magnetic field, $B$, at a small $B_\parallel$. The decrease of the spin spitting is stronger at larger $B_\perp$.  The total magnetic field, $B_{max}$, corresponding to the resistance maximum at different $B_\perp$ is shown in the insert to Fig.\ref{qpmr3}(a).  At high $B_\perp$ the $B_{max}$ is proportional to $B_\perp$ that corresponds to the angle $\alpha_0$=62$^0$.  The position of the maximum agrees, therefore, with the  $B/B_\perp$ scaling.  

The observed  behavior is  compatible with the following relation  between an effective spin slitting $\Delta_{spin}$ and magnetic fields: 
\begin{equation}
\Delta_{spin}=\mu \vert g \vert B+\Delta_\perp,  \Delta_\perp=\beta \hbar \omega_c
\label{zeeman}
\end{equation}
where $\beta<$0.  The parameter $\Delta_\perp$ describes the  additional  contribution of the perpendicular magnetic field to the spin splitting. At the resistance maximum  $\Delta_{spin}$=0 yielding $\beta=-m\vert g \vert\cos\alpha_0/(2m_0)\approx-$0.016$\vert g\vert$.

The structure of the effective spin splitting in Eq.(\ref{zeeman}) is similar to the one used for 2D electron systems.\cite{leadley1998,zudov2012} In particular in Eq.(10) of Ref.[8]: $\Delta_{spin}=\mu g B+\gamma \hbar \omega_c$  the second term  proportional to $B_\perp$ is the  contribution from electron-electron interaction.\cite{ando1982,aleiner1995}  The important difference is, however,  that the sign of the second  term, $\gamma \hbar \omega_c$ is opposite to the sign of the term $\Delta_\perp$ in Eq.(\ref{zeeman}).  Furthermore the magnitude  of the $\beta$ is  an order of magnitude smaller the $\gamma \approx$0.2. The origin of these maxima requires  further investigations.

At $\Delta_Z/\Delta_C>$ 1/2 the angular evolution of the QPMR deviates significantly from the expected behavior.  Instead of periodic oscillations with the parameter $\Delta_Z/\Delta_C$ the resistance demonstrates a weak increase at angles  $\alpha> 86^0$ indicating  that the modulation of the density of states with the energy  does not evolve as expected from Eq.(\ref{QPMR}).  Accounting for the magnetoresistance due to  the in-plane magnetic field (presented in Fig.\ref{qpmr2}(b)) reduces this resistance increase at $\Delta_Z/\Delta_C>$ 1/2 further (not shown).  
\begin{figure}[t]
\vskip 0 cm
\includegraphics[width=\columnwidth]{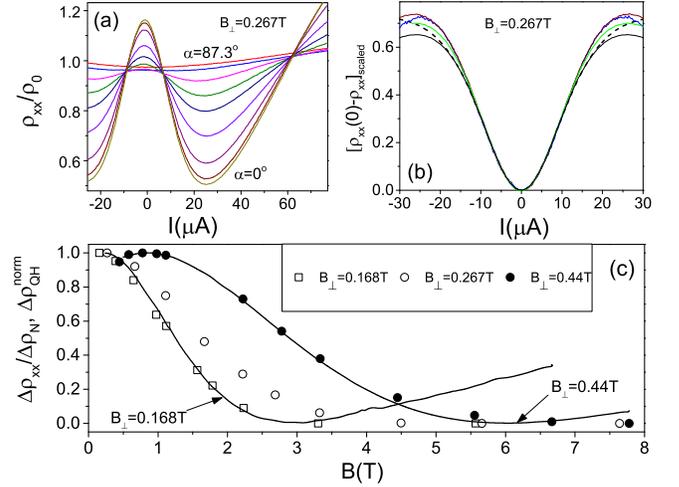}
\caption{(Color online) (a) Quantal heating of 2D electrons at different angles $\alpha$: 0, 66.4, 76.1,80.8, 83.1, 84.3, 85.4, 86.6, 87.3 deg. at $B_\perp$=0.267T; (b) Solid lines are  variations of the  resistance shown in (a), scaled vertically  by factor $k(\alpha)$,  vs the electric current $I$  at several  angles $\alpha$=0, 76.1, 80.8 and 84.3. Dashed line presents fit, which follows  from Eq.(\ref{qh}) for the differential resistance\cite{vitkalov2007}; (c) Solid curves present smoothed  dependencies  of the normalized variations of the resistance  $\Delta \rho_{xx}/\Delta \rho_N$ on the magnetic field $B$.  Symbols present  the normalized magnitude  of the heating induced resistance variation, $\Delta_{QH}^{norm}=k(\alpha)/k_{max}$, obtained at different  magnetic fields $B_\perp$ and $B$ as labeled.}
\label{QH}
\end{figure}

To get a better understanding of the DOS at $\Delta_Z/\Delta_C>$ 1/2 we have conducted measurements of quantal heating. \cite{vitkalov2009,mamani2009,romero2008} Figure \ref{QH}(a) presents dependencies of the normalized differential resistance on the electric current obtained at fixed $B_\perp$ and different total magnetic fields $B=B_\perp/\cos \alpha$. An application of dc current decreases considerably  the differential resistance due to quantal heating. In accordance with  theory the magnitude of the heating induced variation of the conductivity at small perpendicular magnetic fields is proportional to the square of the magnitude of DOS modulations with the energy: $2\delta^2$.\cite{dmitriev2005} Using Eq.(\ref{dos}) for the DOS and Eq.(\ref{cond2}) for the conductivity one can find the effect of quantal heating on the conductivity in a tilted magnetic field following  the case corresponding to  $\alpha$=0$^0$ and considering the inelastic relaxation in the $\tau$-approximation.\cite{dmitriev2005} The variation of the conductivity due to  quantal heating, $\Delta \sigma_{QH}=\sigma(I)-\sigma(0)$,   at $\omega_c\tau_q<$1 is the following:
\begin{equation}
\frac{\Delta\sigma_{QH}}{\sigma_D}=-\delta^2 \left[1+\cos\left(\frac{2\pi \Delta_Z}{\hbar \omega_c}\right)\right]\frac{4Q_{dc}}{1+Q_{dc}}.
\label{qh}
\end{equation}
The term $Q_{dc}=[2\tau_{in}/\tau_{tr}][eER_c]^2[\pi/\omega_c]^2$, where $\tau_{in}$ ($\tau_{tr}$) is inelastic (transport) time,  $R_c$ is cyclotron radius and $E\sim I$ is the electric (Hall) field.\cite{dmitriev2005,vitkalov2007}   Eq.(\ref{qh}) follows from  Eq.(15)  of Ref.[35] if one substitutes  the factor $2\delta^2$   by $\delta^2(1+cos(2\pi\Delta_Z/(\hbar \omega_c))$.  Eq.(\ref{qh}) indicates that the magnitude of the conductivity variation at different angles depends on the factor  $[1+\cos(2\pi \Delta_Z/(\hbar \omega_c)]$, which is identical to the one describing QPMR magnitude (see Eq.(\ref{QPMR})). On the other hand the factor, $Q_{dc}/(1+Q_{dc})$, describing  variations of the resistance with the electric field (current) does not depend on the angle $\alpha$. This means that the shape of the current dependence of the resistance is expected to be the same at different angles, while the overall magnitude of the resistance variations should depend of the angle.

Figure \ref{QH}(b) demonstrates that the heating induced  resistance variations, $\rho_{xx}(0)-\rho_{xx}(I)$,  at different angles $\alpha$ are indeed proportional to each other and to the one expected from Eq.(\ref{qh}).\cite{vitkalov2007}  To reveal the proportionality the  curves, shown in Fig.\ref{QH}(a), are scaled vertically to follow the same dependence on the applied current, $I$. At high currents the dependences deviate from the theory due to  other mechanisms of nonlinearity.\cite{vitkalov2009}  

The normalized magnitude of the heating induced resistance variation $\Delta_{QH}^{norm}=k(\alpha)/k_{max}$ are shown in Fig.\ref{QH}(c) at different $B$ and $B_\perp$.  Here $k(\alpha)$  is the  reciprocal scaling coefficients  of the curves in  Fig.\ref{QH}(b) and  $k_{max}$ is the maximum value of $k$.  At $\Delta_Z/\Delta_C<$ 1/2 the heating induced resistance variations follow the QPMR magnitude in agreement with Eq.(\ref{QPMR}) and Eq.(\ref{qh}) and, thus, correlate with the angular variations of the SdH amplitude. The later  is in agreement with  previous observations.\cite{romero2008}  At $\Delta_Z/\Delta_C>$ 1/2  the heating induced resistance variations are absent, indicating the absence of oscillations of the DOS in this regime. On another hand at $\Delta_Z/\Delta_C<$1/2 the angular evolution of  SdH oscillations, QPMR and quantal heating indicates quantization of the electron spectrum demonstrating  the electron lifetime, $\tau_q\approx$4 ps independent on the angle $\alpha$.   

The results show, thus, a rather $abrupt$  transition of the quantized electron spectrum  at  $\Delta_Z/\Delta_C<$1/2 to an uniform,  energy independent DOS at $\Delta_Z/\Delta_C>$1/2.  Both these results and  investigations of the angular evolution of SdH oscillations\cite{leadley1998,zudov2012}, thus, do not support the proposal  of a gradual decrease  of the quantum scattering time with the  in-plane magnetic field.\cite{zudov2011,bogan2012}  The observed quenching of  MIRO in tilted magnetic fields\cite{mani2005,du2006,bogan2012} also indicates a  modification of the electron spectrum, which happens, however,  at  smaller angles $\alpha<\alpha_c$. This suggests that the transition to an energy independent DOS in the high temperature regime, $kT \gg \hbar \omega_c$,  may depend not only on the ratio between Zeeman and cyclotron energies but also on some other parameters such as electron density, disorder\cite{fogler1995}  and/or the width of the quantum well.

The  angular evolution of  QPMR indicates  significant spin mixing.  This spin mixing suggests an important role  of the spin orbit coupling in  electron transport at high filling factors. The importance of  spin orbit interaction for the quantized spectrum  increases at small  magnetic fields since the strength of this interaction  is independent  of magnetic field.\cite{rashba1960,bychkov1984}     The observed  absence  of QPMR and quantal heating at $\Delta_Z/\Delta_C>$1/2 suggests a transition of the quantized electron orbital motion and the independent periodic spin evolution   to a stochastic spin-orbital dynamics  when energy (period) of the spin  evolution is compatible with the energy (period) of the orbital motion. Below we  evaluate the effect of  spin-orbit interaction on  spin mixing in the studied system.

\subsection{Spin-Orbit Interaction and Quantum Positive Magnetoresistance.}   

Spin-orbit coupling in  quantum wells and heterojunctions has been discussed  in literature\cite{dyakonov_book}. In particular the significant deviation of the g-factor obtained in electrically detected ESR from the bulk GaAs value\cite{stein1983} has been attributed to  spin-orbit effects.\cite{bychkov1984} The spin-orbit interaction leads to  positive quantum corrections to conductivity of disordered 2D conductors.\cite{hikami1980,pikus1994,macdonald2014} In GaAs heterojunctions the effect of spin-orbit interaction  on  quantum corrections to the conductivity has been investigated.\cite{lindelof1997,marcus2003} 

We consider that the spin mixing leading to QPMR is due to  impurity scattering between different $s$-sectors of the Hamiltonian (\ref{SystemHam}) containing a spin-orbit interaction. To evaluate the spin mixing we first find  the electron spectrum,  then compute numerically  matrix elements of the impurity induced  transitions both within an $s$-sector and between different $s$-sectors and compare them.       

We consider a 2DEG in the x-y plane placed in a tilted magnetic field and  affected by  Rashba spin-orbit term. \cite{rashba1960,bychkov1984,zhang2009}.    The in-plane component  of the magnetic field is chosen to be along the $x$-direction yielding ${\bf B}=(B_\parallel,0,B_\perp)$. The Hamiltonian of the system can be written in the following form:

\begin{equation}
\begin{split}
H=&\frac{1}{2m}\left({\bf p}+\frac{e}{c}{\bf A}\right)^2+\frac{\lambda}{\hbar}\hat{{\bf z}}\cdot\left[\left({\bf p}+\frac{e}{c}{\bf A}\right)\times{\bf \sigma}\right]  \\
&+\frac{1}{2}\mu_B g B_\perp \sigma_z+\frac{1}{2}\mu_B g B_\parallel \sigma_{x} 
\end{split}
\label{SystemHam}
\end{equation} 
, where $m, -e$ and $\lambda$ are the electron mass, charge and  spin-orbit coupling constant, respectively and  $\sigma_i$ are the Pauli matrices.  We employ Landau gauge $\mathbf{A}=-yB\hat{\mathbf{x}}$. In that case the Hamiltonian does not contain $x$ variable and the momentum in $x$-direction $p_{x}=\hbar k$ is a conserved quantity. 

As was noted previously \cite{zhang2009} at angle $\alpha=0$ the problem can be solved analytically yielding the following energy spectrum\cite{rashba1960,bychkov1984}
\begin{equation}
E_{n,s}=\hbar\omega_{c}\left(n+\frac{s}{2}\sqrt{(1-g_s)^2+8\eta^2n}\right),
\label{SysEnergy}
\end{equation}
where $\eta=\lambda ml_{B}/\hbar^2$ and $g_s=gm/2m_{0}$. Here $l_{B}=\sqrt{\hbar /eB_\perp}$ is the magnetic length. In Eq.(\ref{SysEnergy}) $s=1$ for $n=0$ and $s=\pm1$ for $n>0$. We note that at $\lambda$=2.5 meV$\cdot$nm  obtained from an analysis of the ESR spectrum in GaAs heterojunctions\cite{stein1983,bychkov1984} the spin-orbital term, $8\eta^2 n$, provides a significant  contribution to the gap between different $s$-sectors in Eq.(\ref{SysEnergy}) at the high filling factors ($n\sim$30) relevant to the experiments.

The corresponding eigenfunctions have the following form
\begin{equation}
\psi_{n,k,s}(x,y)=\cos\theta_{n,s}\chi_{n,k,+1}+ i\sin\theta_{n,s}\chi_{n-1,k,-1} 
\label{SysWaveFunc}
\end{equation}
where  $\theta_{0,1}$=0 and for $n>0$, $\tan\theta_{n,s}=-u_{n}+s\sqrt{u_{n}^2+1}$ and $u_{n}=(1-g_s)/(\eta\sqrt{8n})$. Functions $\chi_{n,k,\sigma}=\phi_{n,k}|\sigma\rangle$ present the eigenfunctions of the Hamiltonian (\ref{SystemHam}) at $\lambda=0$ and $B_\parallel=0$, where $\phi_{n,k}$ are the Landau level  eigenfunctions  and $|\sigma\rangle$ is the eigenstate of the spin operator $\sigma_{z}$ with eigenvalues $\sigma=\pm1$.  Each eigenstate $\psi_{n,k,s}$ has the degeneracy $N_\phi=L_{x}L_{y}eB/(hc)$ related to $N_\phi$ values of $k$, where $L_{x}$ and $L_{y}$ are the system sizes in $x$ and $y$ direction, respectively. 

In a tilted magnetic field, $\alpha>0$,  the problem can be solved numerically \cite{zhang2009}.  An application of the in-plane magnetic field, $B_\parallel$, induces transitions between states $\psi_{n,k,s}$ with different index $s$ (between different $s$-sectors). Using  functions $\psi_{n,k,s}$ as the basis set, one can present the Hamiltonian in  matrix form\cite{mayer2016}.  The matrix contains four  matrix blocks: $\hat H=(\hat E^+,\hat T; \hat T^*, \hat E^-)$, where the semicolon separates rows.  The diagonal matrices $\hat E^+$  and $\hat E^-$  represent energy of the $s$-sectors with  $s=$1  and $s=$-1, respectively,  in different orbital states $n$ following Eq.(\ref{SysEnergy}):    
\begin{equation}
\begin{split}
&E^{+}_{nm}=\delta_{nm}\hbar\omega_{c}\left((n-1) + \frac{1}{2}\sqrt{(1-g_s)^2+8\eta^2(n-1)}\right)\\
&E^{-}_{nm}=\delta_{nm}\hbar\omega_{c}\left(n - \frac{1}{2}\sqrt{(1-g_s)^2+8\eta^2 n}\right)
\end{split}
\label{diag}
\end{equation}
where  indexes $n$=1,2...$N_{max}$ and $m$=1,2...$N_{max}$ numerate rows and columns of the matrix correspondingly.  In  numerical computations the maximum number $N_{max}$ is chosen  to be about twice larger than the orbital number $N_F$ corresponding to Fermi energy $E_F$. Further increase of $N_{max}$ show a very small (within 1\%) deviation from the dependencies obtained at $N_{max}\approx 2N_F$. 
 \begin{figure}[t]
\includegraphics[width=\columnwidth]{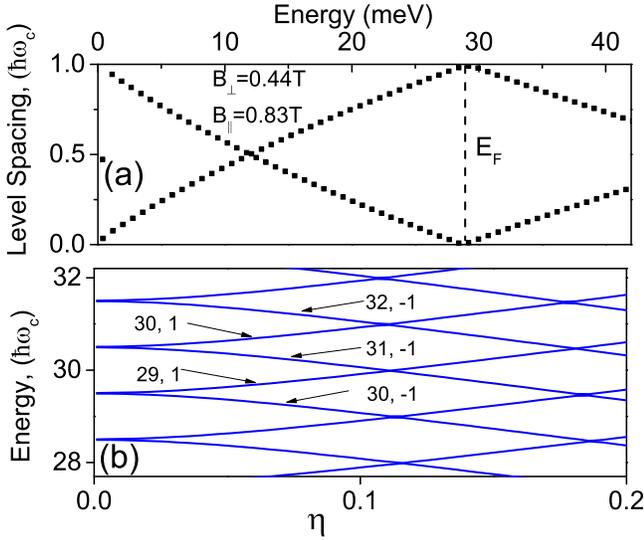}
\caption{(a) Level spacing $\delta E_l=E_{l+1}-E_l$ in the energy spectrum of electrons in  $B_\perp$=0.44 T and $B_\parallel$=0.83 T  at spin-orbit coupling $\lambda$=2.95 meV$\cdot$nm and $g$=-0.44 . 
(b)   Dependence of the energy of the quantum states in the vicinity of Fermi energy  on the spin-orbit coupling parameter $\eta$ at $B_\perp$=0.44T and $\alpha$=0$^0$.  Labels show quantum indexes of the levels according to Eq.(\ref{SysEnergy}).
}
\label{spectrum}
\end{figure} 

The corresponding matrix elements of the off-diagonal matrix $\hat T$ are the following:   
\begin{equation}
T_{nm}= i \delta_{nm}\frac{\mu_B g B_\parallel}{2} \cos\theta_{n-1,1}\sin\theta_{m,-1}
\label{Tmn}
\end{equation}

The Hamiltonian $\hat H$ is diagonalized numerically at different magnetic fields $B_\perp$ and $B_\parallel$. To analyze the spectrum the obtained eigenvalues of the Hamiltonian are numerated in ascending order using positive integer index $l$=1,2.... The electron transport depends on the distribution of the quantum levels in the interval $kT$ near the Fermi energy $E_F$\cite{ziman}.   Below we focus on this part of the spectrum.

Figure \ref{spectrum} presents the difference between energies of $l+1$-th and $l$-th quantum levels of the  electron spectrum. Each symbol represents a particular level spacing normalized to the cyclotron energy: $\delta E_l/\hbar \omega_c=(E_{l+1}-E_l)/\hbar\omega_c$.\cite{mayer2016}.  
Figure \ref{spectrum}(a) presents the normalized level spacing at  spin-orbit coupling $\lambda$=2.95 meV$\cdot$nm and $g$=-0.44 obtained in  $B_\perp$=0.44 T and $B_\parallel$=0.83 T. These magnetic fields correspond to the QPMR maximum  shown in Fig.\ref{qpmr3}. At these conditions the two nearest  quantum levels  coincide in the vicinity of Fermi energy, $E_F$,  yielding the level splitting $\Delta^*$=0.  Fig.\ref{spectrum}(b) presents the  energy of the quantum levels in the vicinity of the Fermi energy vs the strength of the spin-orbit coupling characterized  by the coefficient $\eta$ at $B_\perp$=0.44 T and $\alpha$=0$^0$.  Near $\eta \approx$0.11 levels of two different $s$-sectors intersect opening a channel for the impurity scattering between  $s$-sectors.  Below we evaluate the rate of these transitions for the crossing of the level with quantum numbers $n_0$=30, $s$=1 and the one with $n_0+2$=32, $s$=-1 and investigate the relation of the scattering matrix elements inside the same $s$-sector and between different $s$-sectors.  
\begin{figure}[t]
\includegraphics[width=\columnwidth]{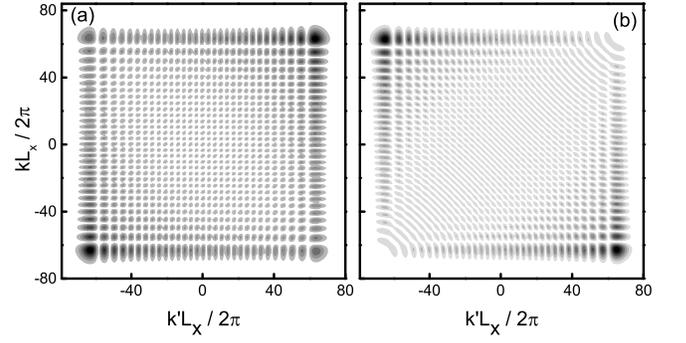}
\caption{The contour plots of $\left|\langle n_{0},1,k|V(x,y)|n_{0},1,k^\prime\rangle\right|$  (a) and $\left|\langle n_{0},1,k|V(x,y)|n_{0}+2,-1,k^\prime\rangle\right|$ (b) for different values of $k$ and $k^\prime$ for  Gaussian impurity potential with parameters  $V_0=0.1\hbar\omega_{c}$, $a=6.6\cdot10^{-4}l_{B}$ and $\eta\approx0.11$. }
\label{s-scattering}
\end{figure} 

We approximate the impurity potential  by  Gaussian function located at $(0,0)$ point:
\begin{equation}
V(x,y)=V_{0}\exp\left[-\frac{x^2+y^2}{2a^2}\right],
\label{ImpPotential}
\end{equation}
where $V_{0}$ is the amplitude of the impurity potential and $a$ defines it's width. For very narrow impurity potential Eq.(\ref{ImpPotential}) can be reduced to a Delta function  $V_{D}(x,y)=2\pi V_{0}a^2\delta(x)\delta(y)$. In this  case at $\alpha$=0$^0$ the matrix elements can be written explicitly:

\begin{align}
\langle n,s,k&|V_{D}(x,y)|n^\prime,s^\prime,k^\prime\rangle=2\pi V_{0}a^2\times \nonumber \\
&\bigg[\sin\theta_{n,s}\sin\theta_{n^\prime,s^\prime}\phi_{n-1,k}(0,0)\phi_{n^\prime-1,k^\prime}(0,0)+ \nonumber \\
&\qquad\cos\theta_{n,s}\cos\theta_{n^\prime,s^\prime}\phi_{n,k}(0,0)\phi_{n^\prime,k^\prime}(0,0)\bigg],
\label{delta}
\end{align}
whereas for general case they should be computed numerically. 

Below we compute the matrix elements inside the same $s$-sector $\langle n_{0},1,k|V(x,y)|n_{0},1,k^\prime\rangle$ and between different sectors $\langle n_{0},1,k|V(x,y)|n_{0}+2,-1,k^\prime\rangle$ at the value  $\eta\approx$0.11 and compare the  magnitudes of these two matrix elements.  In  calculations the size of the system in the $y$-direction is $L_{y}=6R_{c}$, where $R_c=\sqrt{2n_{0}+1}l_{B}$ is the cyclotron radius. 

Fig.\ref{s-scattering}(a) shows  a contour plot of the magnitude of the matrix element $\left|\langle n_{0},1,k|V(x,y)|n_{0},1,k^\prime\rangle\right|$  within the same $s$-sector while  Fig.\ref{s-scattering}(b) shows the magnitude of the impurity scattering between different $s$-sectors $\left|\langle n_{0},1,k|V(x,y)|n_{0}+2,-1,k^\prime\rangle\right|$  for different values of $k$ and $k^\prime$. The impurity parameters are $V_0=0.1\hbar\omega_{c}$ and $a=6.6\cdot10^{-4}l_{B}$. Fig.\ref{s-scattering}(a) demonstrates that for scattering within the same sector  both forward  scattering ($k=k^\prime$) and backscattering ($k=-k^\prime$) are substantial, although  forward scattering is somewhat  stronger than backscattering. In contrast, in  transitions between different $s$-sectors  backscattering plays the major role while forward scattering is strongly suppressed. The average of  the squares of  scattering amplitudes  are found to be of the same order:  $1.22\cdot10^{-19}(\hbar\omega_c)^2$ within  the same $s$-sector and $0.72\cdot10^{-19}(\hbar\omega_c)^2$  between  different $s$-sectors. Thus  due to  backscattering  the impurity scattering between different $s$-sectors is comparable  with that within the same $s$-sector. We note that the studied systems demonstrate a significant magnitude of  impurity backscattering.\cite{du2002,glazman2007,mayer2016b}

\begin{figure}[t]
\includegraphics[width=8.5cm]{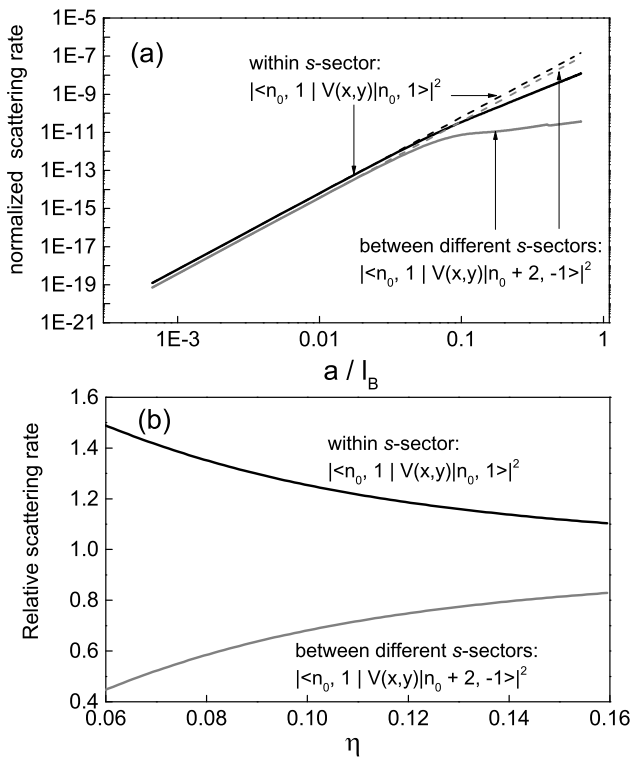}
\caption{ (a) The dependence of the square of matrix elements on impurity width parameter $a$ both for Gaussian $V(x,y)$ and Delta $V_{D}(x,y)$ function potential cases for $\eta\approx0.11$. (b) The dependence of the square of matrix elements on the value of spin-orbit interaction parameter $\eta$ for Gaussian impurity with the width parameter  $a=6.6\cdot10^{-4}l_{B}$. The impurity amplitude is  $V_0=0.1\hbar\omega_{c}$ for both figures and the amplitudes of matrix elements are averaged over all $k$ and $k^\prime$ values.}
\label{s2} 
\end{figure}

Fig.\ref{s2}(a) presents the dependence of the averaged square of the matrix elements on the shape of the impurity potential $V(x,y)$ at $\eta\approx0.11$. The average is  over all $k$ and $k^\prime$ values. This figure shows that at $a<0.05l_B$ both Gaussian  and Delta function potentials  provide nearly identical scattering both within the $s$-sector and between different $s$-sectors and  the scattering magnitude is proportional to the cross-section of the impurity potential $a^2$.  At  higher magnetic fields  $a>0.05l_{B}$  the scattering on the Gaussian potential deviates from the $a^2$ dependence. More importantly the figure shows that at $a>0.05l_{B}$ the  impurity potential cannot provide significant  scattering between different $s$-sectors. Thus  the scattering between different $s$-sectors is effective at relatively small magnetic fields (high filling factors) and/or for  sharp impurities.  At the upper limit of the perpendicular magnetic fields used in this study, $B_\perp \approx$1 T,    the magnetic length $l_{B} \approx$25 nm and  for  impurities with  size  $a$ less 1nm  backscattering is effective and  leads to  the strong spin mixing at $B_\perp<$1T.   The size, $a<$1 nm,   is reasonable for neutral impurities in a solid. 

Due to the impurity scattering  quantum levels are broadened and the elastic transitions may occur in an interval of the energies when two  levels overlap. Thus the scattering may exist in an interval of the parameter $\eta$. Figure \ref{s2}(b) presents the  dependence of the averaged square of matrix elements on the parameter $\eta$. The $\eta$ is varied in the range, where the energy of the system changes by about $0.6\hbar\omega_{c}$. It accounts, thus, for a significant broadening of  quantum levels.  The figure shows that the amplitudes of both the intra-sector and inter-sectors scattering  are quite comparable in the broad range of $\eta$ and the difference decreases with the $\eta$ increase.   The increase of the scattering between different $s$-sectors is  related to the fact that at $\eta=0$  different sectors correspond to eigenstates with different $z$ components of the electron spin. These states cannot be coupled by the impurity scattering unless a magnetic impurity is involved (see Eq.(\ref{delta})). Due to the fact that the majority of the impurities in the studied systems are non-magnetic  the scattering between different sectors is completely mediated by the spin-orbit interaction and increases with the increase of  the spin-orbit coupling.

The presented estimations of the impurity scattering  in the presence of spin-orbit interaction indicate  that in the range of physical parameters relevant to presented experiments the scattering between different $s$-sectors is comparable with the scattering within the same $s$-sector. This leads to  strong  spin mixing in the studied systems and, thus, support the assumption used for Eq.(\ref{cond}). 

\section{Conclusion}

 Quantum positive magnetoresistance (QPMR)   of 2D electrons is studied at different angles $\alpha$ between the magnetic field and the normal to the 2D layer.  The magnitude of  QPMR varies significantly with  the magnetic field tilt. The angular evolution of QPMR correlate strongly with angular variations of the amplitude of SdH resistance oscillations indicating that the Zeeman spin splitting, $\Delta_Z$,  enhanced by electron-electron interaction, is the dominant mechanism leading to the QPMR reduction. Surprisingly no quantization of the electron spectrum is detected when  the Zeeman energy exceeds the half of the cyclotron energy suggesting a  transformation of the electron dynamics in the high temperature regime at $kT \gg\Delta_Z>\hbar\omega_c/2$ .

In contrast to SdH oscillations the angular evolution of QPMR implies substantial  mixing between spin subbands. A spin mixing have been detected in other 2D electrons systems.\cite{vitkalov2000,vitkalov_hall}   Although the origin of the spin mixing remains puzzling investigations  indicate, that the spin-orbit interaction  may lead to a significant spin mixing via impurity scattering in the studied system.

This work was supported by the National Science Foundation (Division of Material Research - 1104503),  the Russian Foundation for Basic Research (Project No. 14-02-01158) and  the Ministry of Education and Science of the Russian Federation. We also acknowledge support from NSF EFRI 1542863 (AG) and (PG).

\end{document}